\documentclass[aps,prb,twocolumn,english,aps,amsmath,amssymb,showpacs,prb,superscriptaddress]{revtex4-1} \usepackage[T1]{fontenc} \usepackage[latin9]{inputenc} 
\usepackage{float}
\usepackage{amstext} 
\usepackage{graphicx}
\usepackage{color}

\makeatletter

\@ifundefined{textcolor}{}
{
 \definecolor{BLACK}{gray}{0}
 \definecolor{WHITE}{gray}{1}
 \definecolor{RED}{rgb}{1,0,0}
 \definecolor{GREEN}{rgb}{0,1,0}
 \definecolor{BLUE}{rgb}{0,0,1}
 \definecolor{CYAN}{cmyk}{1,0,0,0}
 \definecolor{MAGENTA}{cmyk}{0,1,0,0}
 \definecolor{YELLOW}{cmyk}{0,0,1,0}
}

\makeatother
\usepackage{babel} \begin{document}

\title{Electronic structure of Ce$_{2}$RhIn$_{8}$ 2D heavy Fermion system studied by angle resolved photoemission spectroscopy}

\author{Rui Jiang}
\affiliation{Ames Laboratory and Department of Physics and Astronomy, Iowa State University, Ames, Iowa, 50011, USA} 
\affiliation{Department of Physics and Astronomy, Iowa State
University, Ames, IA 50011, USA}

\author{Daixing Mou}
\affiliation{Ames Laboratory and Department of Physics and Astronomy, Iowa State University, Ames, Iowa, 50011, USA} 
\affiliation{Department of Physics and Astronomy, Iowa State
University, Ames, IA 50011, USA}

\author{Chang Liu}
\affiliation{South University of Science and Technology of China, Shenzhen, China 518055}

\author{Xin Zhao}
\affiliation{Ames Laboratory and Department of Physics and Astronomy, Iowa State University, Ames, Iowa, 50011, USA}
\affiliation{Department of Physics and Astronomy, Iowa State
University, Ames, IA 50011, USA}

\author{Yongxin Yao}
\affiliation{Ames Laboratory and Department of Physics and Astronomy, Iowa State University, Ames, Iowa, 50011, USA}
\affiliation{Department of Physics and Astronomy, Iowa State
University, Ames, IA 50011, USA}

\author{Hyejin Ryu}
\affiliation{Condensed Matter Physics and Materials Science Department, Brookhaven National Laboratory, Upton, NY, 11973, USA}
\affiliation{Department of Physics and Astronomy, Stony Brook University, Stony Brook, New York 11794-3800, USA}

\author{C. Petrovic}
\affiliation{Condensed Matter Physics and Materials Science Department, Brookhaven National Laboratory, Upton, NY, 11973, USA}
\affiliation{Department of Physics and Astronomy, Stony Brook University, Stony Brook, New York 11794-3800, USA}

\author{Kai-Ming Ho}
\affiliation{Ames Laboratory and Department of Physics and Astronomy, Iowa State University, Ames, Iowa, 50011, USA}
\affiliation{Department of Physics and Astronomy, Iowa State
University, Ames, IA 50011, USA}

\author{Adam Kaminski}
\affiliation{Ames Laboratory and Department of Physics and Astronomy, Iowa State University, Ames, Iowa, 50011, USA} \affiliation{Department of Physics and Astronomy, Iowa State
University, Ames, IA 50011, USA}

\date{\today} 

\begin{abstract} We use angle-resolved photoemission spectroscopy (ARPES) to study the 2D heavy fermion superconductor, Ce$_2$RhIn$_8$. The Fermi surface is rather complicated and consists of several hole and electron pockets with one of the sheets displaying strong nesting properties with a q-vector of (0.32, 0.32) $\pi$/a We do not observe k$_z$ dispersion of the Fermi sheets, which is consistent with the expected 2D character of the electronic structure. Comparison of the ARPES data to band structure calculations suggest that a localized picture of the f-electrons works best. While there is some agreement in the overall band dispersion and location of the Fermi sheets, the model does not reproduce all observed bands and is not completely accurate for those it does. Our data paves way for improving the band structure calculations and the general understanding of the transport and thermodynamical properties of this material. 
\end{abstract} 

\maketitle

\section{Introduction}

Heavy fermions, first discovered in 1975\cite{Andres:1975ia}, are some of the most fascinating materials in condensed matter physics\cite{RevModPhys.56.755}. The name originates from the enhanced effective mass of the quasi-particles, which can be two or three orders of magnitude higher than in a normal metal, while the functional form of the resistivity remains the same as a Fermi liquid. The behavior of this system is dominated by the 4f and 5f electrons and arises from a competition between the Kondo effect and the Ruderman-Kittel-Kasuya-Yosida (RKKY) interaction. Due to the complexity of this interaction, many marvelous phenomena are seen in heavy fermion compounds, such as unconventional superconductivity\cite{PhysRevLett.43.1892}, quantum criticality\cite{Coleman:2005gb}, a possible topological insulator\cite{Dzero:2010dj} and many others. Since the electrons in heavy fermion materials are an important test-bed for understanding the interplay between magnetic and electronic quantum fluctuations, measurements of the electronic structure are a crucial step for further research\cite{Coleman:2006ua}. Band structure calculations for heavy fermion materials are difficult because of the complex behavior of the f-electrons of the rare earth elements. An additional complication is the difficulty in obtaining  experimental data from techniques such as Angle Resolved Photoemission Spectroscopy that directly measure the band structure and other electronic properties. The 3D character of the band structure found in the vast majority of heavy fermion materials makes investigation with ARPES difficult because of final state broadening effects \cite{Hansen:1998gb}, and projection effects of 3D bands due to limited k$_z$ selectivity, etc. One notable exception is Ce$_{2}$RhIn$_{8}$. This is a highly layered material thought to have a quasi two dimensional electronic structure\cite{Ueda:2004em}. This opens up the possibility for a detailed ARPES study of its band structure and other electronic properties and the application of this powerful technique to uncover spectroscopic features. Knowledge of the electronic properties such as the location and shape of the Fermi surface sheets, Fermi velocities, band renormalization, scattering rates and interactions with collective excitations can then be used to model the thermodynamic and transport properties.  

Ce$_{2}$RhIn$_{8}$ with lattice parameters a=4.665\AA, c=12.244\AA ~and tetragonal crystal structure\cite{Bao:2001gt} is an antiferromagnetic member of the Ce$_n$MIn$_{3n+2}$ (M=Co, Rh or Ir, n=1, 2 or $\infty$) family of heavy-fermion materials with two other compounds being unconventional superconductors. Since the structure of Ce$_{2}$RhIn$_{8}$ can be viewed as inserting a CeIn$_{3}$ into CeRhIn$_5$, this material is  expected to share some of the properties of both compounds. The enhanced value of the Sommerfeld coefficient\cite{Cornelius:2001ds} ($\sim400~\mathrm{mJ/molCeK^2}$) determined by specific heat measurements is consistent with  the heavy fermion nature of this material. The resistivity curve\cite{Malinowski:2003gq} follows a $\mathrm{ln}(1/T)$ behavior between 55K and 130K as a result of Kondo screening and a single impurity model estimate of the Kondo temperature $T_K$ yields a value of 10K. The slope of the resistivity changes at $T_N=2.8K$ and $T_{LN}=1.65K$ indicating two magnetic transitions. Neutron scattering measurements\cite{Bao:2001gt} performed at 1.6K show the presence of an anti-ferromagnetic state with ordering vector $Q=\left(\frac{1}{2},\frac{1}{2},0\right)$. At this temperature, the magnetic moment of Ce is well screened at 0.55$\mu_B$, compared with 2.35$\mu_B$ per Ce at high temperature (200K). The slope of the resistivity changes more dramatically at $T_{LN}$ compared to $T_N$, and $T_{LN}$ is also more sensitive to pressure with $P_c\sim0.04$GPa, suggesting that the magnetic structure changes from an incommensurate to a commensurate one at $T_{LN}$\cite{Malinowski:2003gq}. The nature of the AFM order seems to fit better with a scenario of local moment ordering rather than SDW ordering\cite{Cornelius:2001ds}. This is because an additional term would be needed to fit the specific heat data in order to account for an anisotropic gap in the SDW state for CeRhIn$_5$. Also, according to conventional models of antiferromagnetic quantum criticality\cite{Millis:1993fy}, a linear decrease of $T_N$ with pressure points to an effective 2D character of the spin-fluctuation spectrum. This favors a Kondo destruction scenario with local moment ordering \cite{Coleman:2012ci}.


Due to the layered structure, the electronic properties of Ce$_{2}$RhIn$_{8}$ are believed to be quasi-2D, which is rarely seen in heavy-fermion superconductors. ARPES data have been previously reported\cite{Raj:2005ej,Souma:2008jo} for this compound but only along high symmetry directions. To the best of our knowledge there have been no reported measurements of the Fermi surface. To better understand the superconductivity and heavy-fermion phenomenon in this material, we examine the Fermi Surface and detailed band dispersion of Ce$_{2}$RhIn$_{8}$ using various photon energies.

\begin{figure}[h!tb] \centering
 \includegraphics[width=\linewidth]{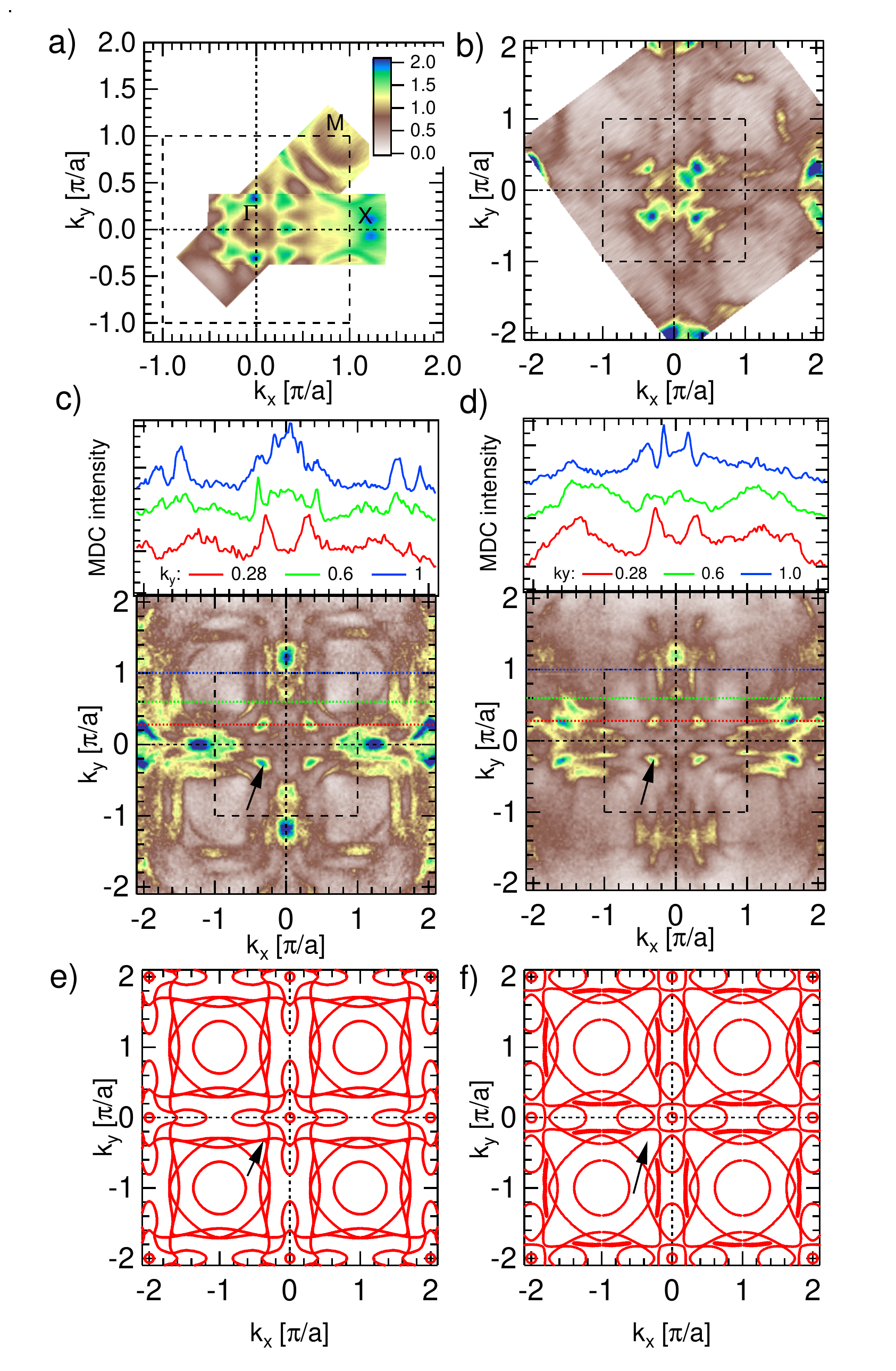}
 \caption{ Fermi Surface measured with (a) a He lamp (21.2eV) at 16K. (b) A synchrotron (SRC, 80eV) at 20K. (c) A synchrotron (ALS, 94eV) at 17K. The top panel shows MDC data at E$_F$ for selected values of k$_y$ marked by color coded lines in the lower panel. (d) synchrotron (ALS, 105eV) at 17K. Black dashed rectangles show the boundary of the first Brillouin zone. Top of the panel shows MDC at E$_F$ for selected values of k$_y$ marked by dashed lines. (e)\&(f) The Fermi surface calculated by DFT with the f electron set as localized and itinerant respectively.}
  \label{Fig1} 
 \end{figure}
 
\section{Methods}

Single crystals were grown using the In flux technique and characterized as described in  Ref. 9. ARPES measurements were performed using the ARPES system at Ames Laboratory and beamline 7.0.1 at the Advanced Light Source (ALS). Samples were cleaved \textit{in situ} yielding shiny, mirror-like surfaces. All ARPES data were taken at T=16K, above the AFM transition temperature (2.8K) but close to the Kondo temperature (10K). The laboratory-based ARPES system consists of a GammaData ultraviolet lamp (21.2eV He I$\alpha$), custom-designed refocusing optics and a Scienta SES2002 electron analyzer. The UV spot size is around 1mm and the energy resolution was set at 10meV. Beamline 7.0.1 is equipped with a Scienta R4000 electron analyzer with energy resolution around 40meV.

First-principles band structure calculations were performed using spin-polarized density functional theory (DFT)\cite{Kohn:1965js} within a generalized-gradient approximation (GGA) with the projector-augmented wave (PAW) method\cite{Blochl:1994dx,Kresse:1999dk} using VASP code\cite{Kresse:1996kl}. The GGA exchange correlation functional parameterized by Perdew, Burke and Ernzerhof (PBE)\cite{Perdew:1996iq} was used. The semi core p-states of Rh, as well as the lower lying d-states of In, were treated as valence states, while the 4f electrons are treated in two ways for comparison; either placed in the core or as valence states. The kinetic energy cutoff was 400 eV and the Monkhorst-PackÕs scheme\cite{Monkhorst:1976cv} was used for Brillouin zone sampling with a k-mesh of 17 x 17 x 7 for the FM state and 12 x 12 x 7 for the AFM state in which case a 2 x 2 x 1 supercell was used. 

\begin{table}[h!tb] \centering
\setlength{\tabcolsep}{2pt}
\caption{Optimized lattice parameters, energies and magnetic moment of the Ce atoms in different magnetic states.}
\label{Ce218_cm}
\begin{tabular}{c|cccc} \hline
 & a(\AA) & c(\AA) & M$_{Ce}$ ($\mu_B$/Ce atom) & E (eV/atom) \\ \hline
AFM & 6.651 & 12.283 & 0.69 & -4.029\\
FM & 4.709 & 12.278 & 0.68 & -4.030 \\
Non-Mag & 4.705 & 12.280 & 0 & -4.027 \\
Expriment\cite{Bao:2001gt} & 4.665 & 12.244 & 0.55 & - \\ \hline
\end{tabular}
\end{table}

Calculated lattice parameters, energies and magnetic moment of the Ce atoms in different magnetic states are listed in table~\ref{Ce218_cm}, together with the experimental results\cite{Bao:2001gt}. The lattice parameters obtained from the GGA calculations are in good agreement with experiment, with an approximate 1\% overestimation. The calculated magnetic moment of the Ce ions also agrees well with the experimental data. From table~\ref{Ce218_cm}, we can say that the magnetically ordered states have slightly lower energy than the non-magnetic state, with the energy of the FM state being  slightly lower by $\sim$1 meV per atom.

\begin{figure}[h!tb] \centering
\includegraphics[width=\linewidth]{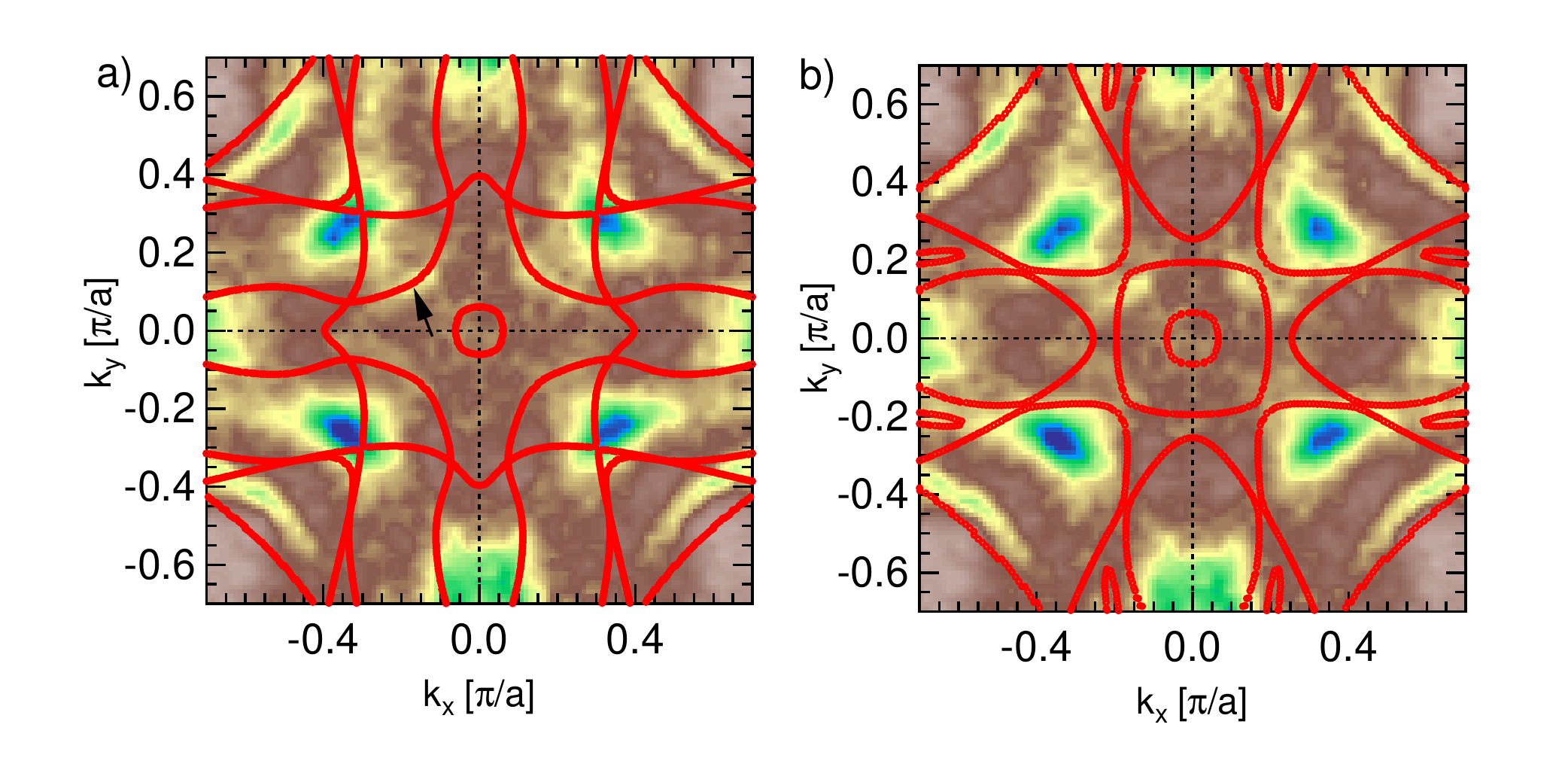}
\caption{Fermi surface image close to $\Gamma$ point measured using 94 eV photon energy with overlay of  Fermi surface calculation using (a) localized or (b) itinerant scenario.}
\label{Fig2} 
\end{figure}

\begin{figure}[h!tb] \centering
\includegraphics[width=\linewidth]{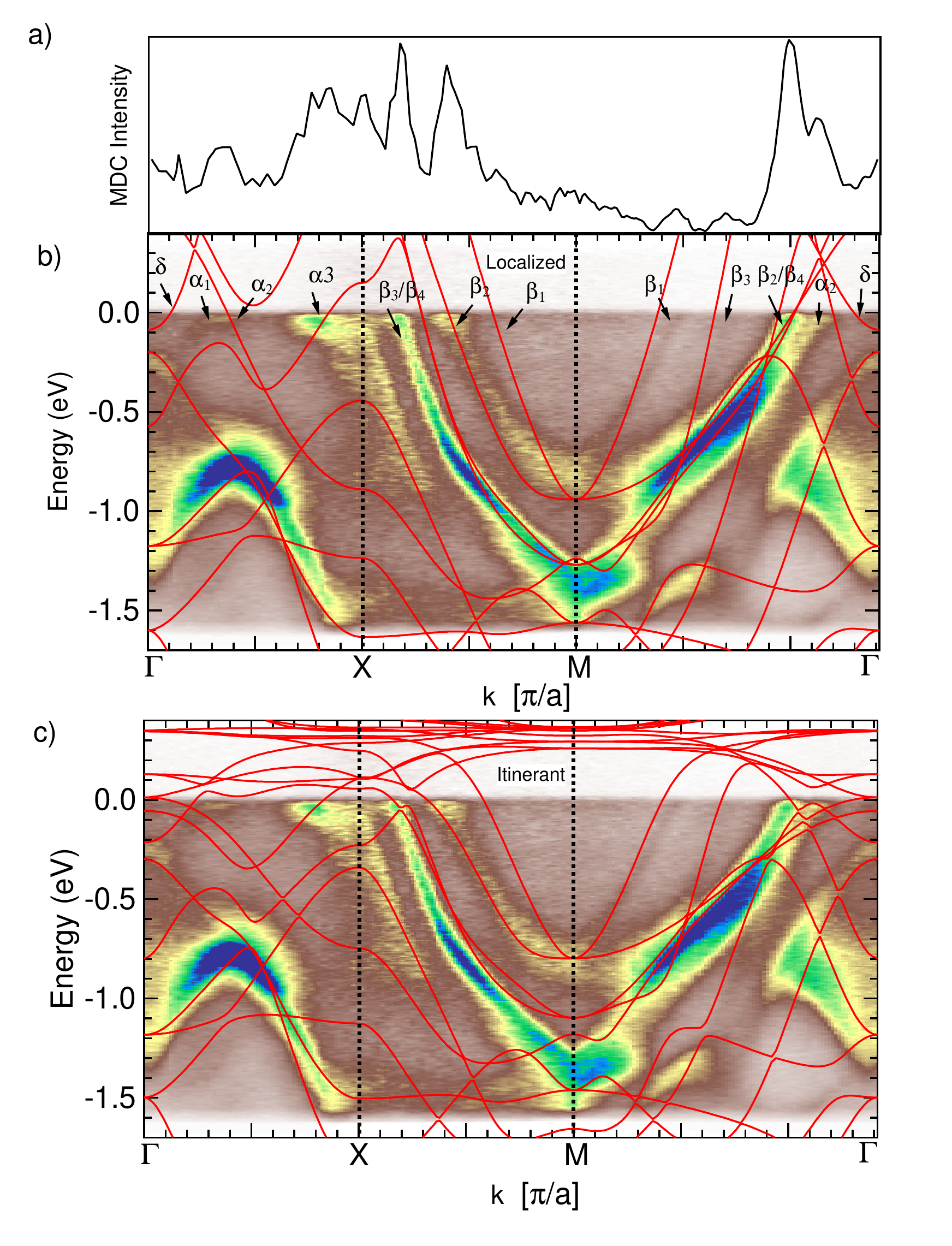}
\caption{Intensity plot along the high symmetry directions overlaid with a localized f-electron band structure calculation. The top panel shows the MDC intensity at E$_F$. Labelled black arrows mark the band crossings in the data.}
\label{Fig3} 
\end{figure}

\section{Results and discussion}

In Fig.~\ref{Fig1} (a)-(d) we plot the Fermi surface of Ce$_{2}$RhIn$_{8}$ obtained at  several photon energies by integrating the photoelectron intensity within $\pm5$meV of the E$_F$. Dark brown, yellow, green and blue areas mark the locations of the Fermi surface sheets. Data in  panel (a) was measured using a laboratory He source with a photon energy of 21.2eV, (b) was measured at SRC using 80eV photons. Data in panels (c) \& (d) was measured at ALS using 94eV and 105eV photons, respectively. Since the cross section of the bands can be quite different for different photon energies and polarizations, performing the measurements with several photon energies can reveal a more complete picture of the Fermi surface. The topology of the Fermi surface of Ce$_{2}$RhIn$_{8}$ is rather complicated and consists of several large electron pockets ($\beta_1$ to $\beta_4$ labeled in Fig. 3b) centered at the M-point, a large and small electron ($\alpha_1$ and $\delta$ labeled in Fig. 3b ) pocket and a hole ($\alpha_2$ labeled in Fig. 3b) pocket near the center of the zone. We compare this data to the calculated Fermi surface for a localized (panel e) and an itinerant (panel f) model of the f-electrons. The magnetic moment for the itinerant scenario was artificially set to zero, since the data was measured at a temperature above $T_N$. The Fermi surface for those two scenarios are very similar around the M-point but quite different around the $\Gamma$-point, where the localized approach fits the experimental data better. Close to $\Gamma$, there is an area of high intensity for all measured photon energies due to one or more Fermi sheets marked by black arrows. In the localized model, this area coincides with the presence of two adjacent bands at E$_F$ (arrow in panel e). The  itinerant model on the other hand does not predict any Fermi sheets in that area (arrow in panel f). This is not surprising since the data was measured at 16K. Although the temperature is comparable with $T_N\approx10K$ and screening of the f electrons by conduction electrons should exist to some extent, it is still much higher than the coherence temperature of 5K and most f-electrons will remain localized. A more detailed comparison is presented in Fig. 2, where we overlay the calculated Fermi surface for both models onto the ARPES data measured at 94 eV. Again the localized model fits the data much better, while the itinerant approach fails to predict the observed location of the Fermi surface sheets.


We now proceed to compare the calculated and measured band dispersions. In Fig. 3 we plot the ARPES intensities along the main symmetry directions measured at 94 eV photon energy. Overlaid on top is the calculated band structure using the itinerant and localized models. The dark brown, yellow, green and blue areas mark the locations of the bands. The localized approach fits the data better near the $\Gamma$ point and there are some overall similarities in the shape and location of some bands, however there are also some striking differences. For example the experimental data shows a dispersing band crossing E$_F$ between $\alpha_3$ and $\beta_3$, which neither calculation predicts. This underlines the need to improve our basic understanding of the electronic structure and hopefully the data presented here will guide that effort. To facilitate this we plot in Fig. 4 several more-detailed intensity plots near the center of the zone plus an off-symmetry set of data showing the shallow band along the cut at k$_y$=0.7 $\pi/a$.

\begin{figure}[h!tb] \centering
 \includegraphics[width=\linewidth]{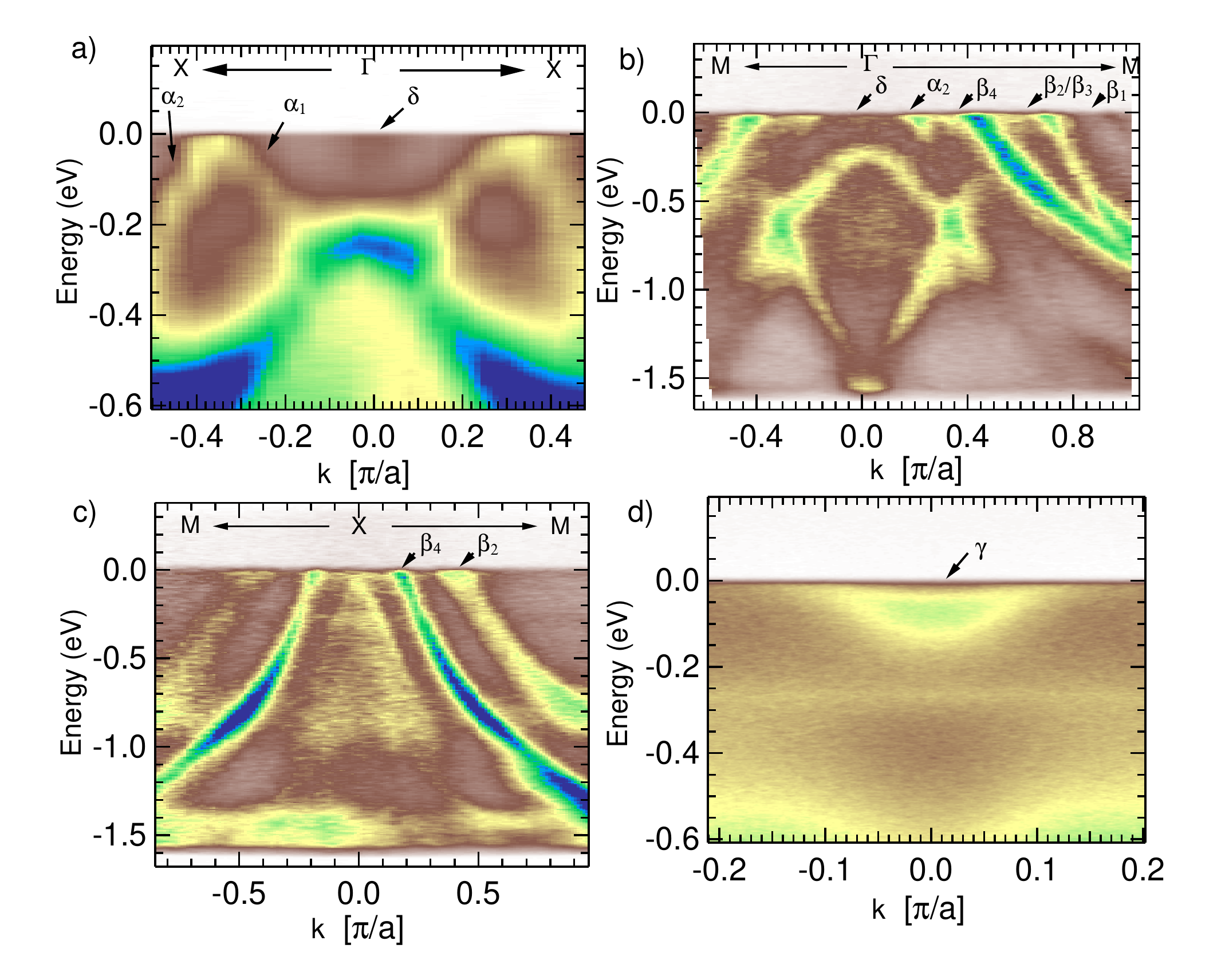}
 \caption{ARPES data along selected cuts showing details of the band structure: a) X-$\Gamma$-X  b) M-$\Gamma$-M c) M-X-M d) cut along k$_y$ direction for k$_x$=0.7 $\pi/a$. } 
 \label{Fig4} 
 \end{figure}

\begin{figure}[h!tb] \centering
 \includegraphics[width=\linewidth]{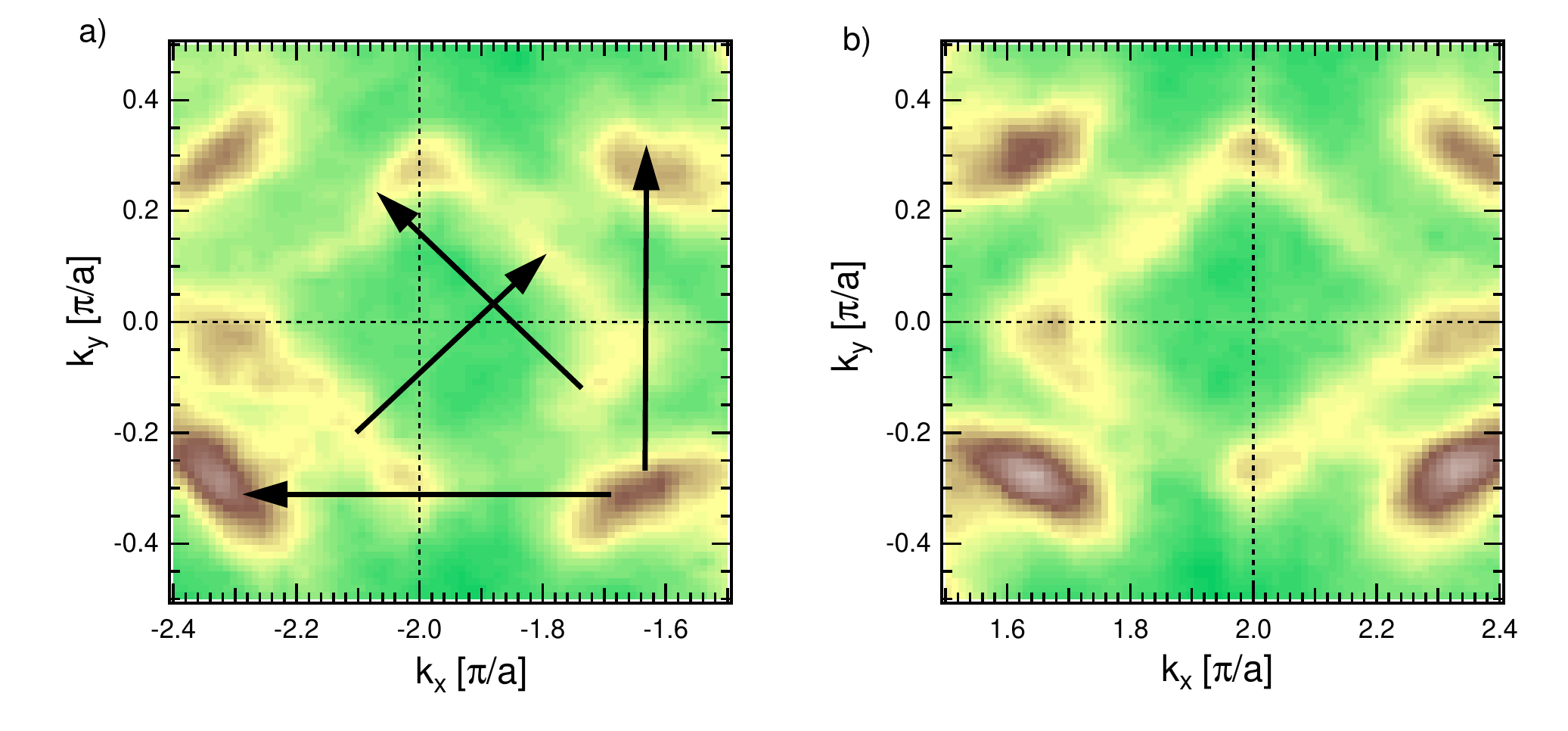}
 \caption{Intensity maps at the E$_f$ near the center of the second Brillouin zone for a) negative  and b) positive k$_x$. Arrows mark a possible nesting vector (0.32, 0.32) $\pi/a$ connecting parallel parts of the Fermi surface.}
  \label{Fig5} 
 \end{figure}

To reveal the nature of the AFM order, one needs to establish the presence of a nesting vector equal to the AFM ordering vector $\mathbf{q}$=(0.5,0.5,0).  The high intensity spot at the corner of bands $\beta_2$ and $\beta_4$ (seen in Fig. 5) corresponds to a slightly larger nesting vector $\sim$(0.6, 0.6, 0) than the ordering vector $\mathbf{q}$. This nesting vector may decrease below the coherence temperature with the injection of f-electrons. Since the Fermi surface of Ce$_{2}$RhIn$_{8}$ contains multiple pockets, when nesting occurs for one of the pockets, a large part of Fermi surface is unaffected. This is consistent with specific heat measurements\cite{Cornelius:2001ds}, where the change of the Sommerfeld coefficient $\gamma$ above and below $T_N$ (from 400 to 370~$\mathrm{mJ/molCeK^2}$) is much smaller than in CeRhIn$_5$ (400 to 56~$\mathrm{mJ/molCeK^2}$), indicating that only a small part ($\sim8\%$) of the Fermi surface becomes gapped below T$_N$. Therefore, the spin density wave scenario cannot be completely ruled out by our measurements.  We also observe a well-nested diamond shape Fermi surface around $\Gamma$ with a nesting vector of (0.32, 0.32, 0) marked by the diagonal arrows. There are however no obvious consequences of this nesting in the neutron inelastic scattering data. It is imperative to study how this part of the Fermi surface evolves as the sample is cooled below the Neel temperature to find out the role these segments play in the low temperature properties. The nearly straight segments of this Fermi surface sheet deviate from the calculation results, which predict more rounded shape. This again underlines need to revise our understanding of band structure and improve the computational model.

\begin{figure}[h!tb] \centering
 \includegraphics[width=\linewidth]{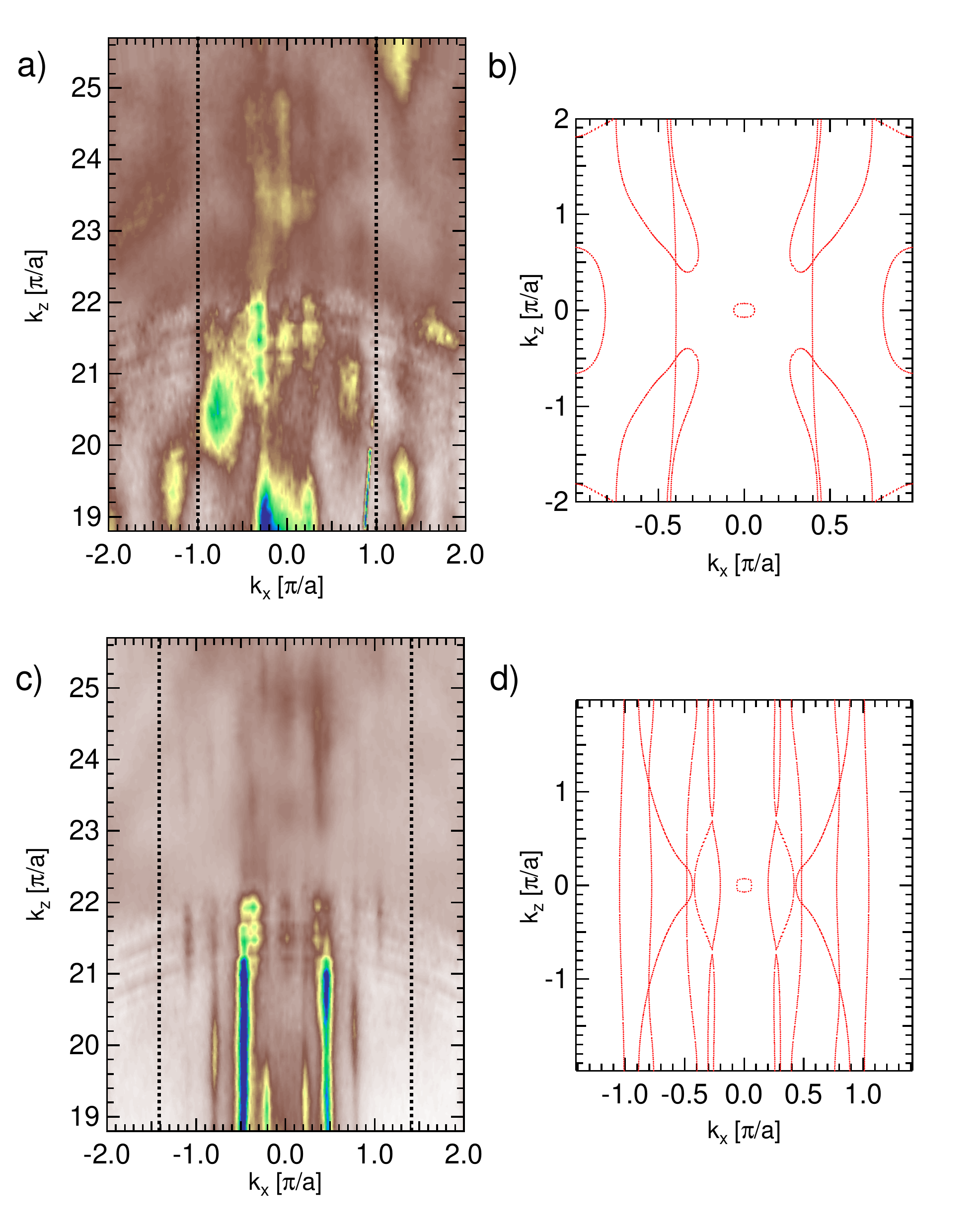}
 \caption{$k_z$ dispersion along (a,b) the $\Gamma$ - X direction and (c,d) the $\Gamma$ - M direction from 80eV to 157eV compared with the localized f-electron calculation. } 
  \label{Fig6} 
 \end{figure}

The Ce$_2$RhIn$_8$ is deemed to be a very important heavy fermion material because indications are that its electronic structure has a 2D character. To date, this was not been directly demonstrated by ARPES measurements because the data in Refs. \cite{Raj:2005ej,Souma:2008jo} was acquired using a single photon energy, which probes only single surface in 3D momentum space. To verify the dimensionality of the electronic structure of this material we present extensive k$_z$ dispersion data measured using photon energies from 80 eV to 157 eV in Fig. 6. Assuming a reasonable value for the inner potential of 14 eV, this corresponds to k$_z$ values from 18.9 $\pi$/c to 25.8 $\pi$/c. The data in panel (a) was measured along $\Gamma$-X and panel (b) shows data along the $\Gamma$-M direction. Even though the intensity of the Fermi sheets changes with photon energy, all observed Fermi sheets present themselves as vertical lines with no observable dispersion. This demonstrates the quasi-2D character of the electronic structure of Ce$_2$RhIn$_8$, which is consistent  with the 2D effective dimensionality of the spin-fluctuation spectrum from the phase diagram\cite{Nicklas:2003cy}. This lack of dispersion in the data prevents the exact determination of the offset for  the k$_z$ values on the vertical axes of Fig. 6a, c. The calculated $k_z$ dispersion is shown in Fig. 6 b, d. The localized model calculations predicts a nearly 2D electronic structure along the the $\Gamma$-M direction. The model predicts some dispersion for the $\Gamma$-X direction which is clearly not observed in the data.

\section{Conclusions}

We used angle-resolved photoemission spectroscopy to measure the electronic properties of Ce$_{2}$RhIn$_{8}$. The lack of a significant $k_{z}$ dispersion confirms the quasi two dimensional nature of the electronic structure. The measured Fermi surface is quite complicated and consists of several hole and electron pockets. By comparing our data with a DFT calculation, we find our results are consistent with a localized picture of the f-electrons. We also report some striking differences between the measured and calculated band dispersion, underlining some basic inadequacies of the calculated approach. The presented data will likely guide the development of new theoretical approaches that better address systems with f-electrons and are applicable to heavy fermion compounds.

\section{Acknowledgments} 

We would like to thank Bruce Harmon for useful discussions. This work was supported by the U.S. Department of Energy, Office of Science, Basic Energy Sciences, Materials Science and Engineering Division.  Ames Laboratory is operated for the U.S. DOE by Iowa State University under contract \# DE-AC02-07CH11358 (ARPES measurements and data analysis). The Advanced Light Source is supported by the Director, Office of Science, Office of Basic Energy Sciences, of the U.S. Department of Energy under Contract No. DE-AC02-05CH11231. Work at Brookhaven was supported by the U.S. DOE under Contract No. DE-AC02-
98CH10886 (sample growth and characterization).

\bibliography{mybib}{}
\end{document}